\documentclass[twocolumn,aps,prd,showpacs,floatfix]{revtex4}

\usepackage{latexsym}
\usepackage{amsmath}
\usepackage{amssymb}
\usepackage{latexsym}
\usepackage{color}
\usepackage{graphicx}
\usepackage{cancel}
\usepackage{bbm}
\usepackage{bm}
\usepackage{maybemath}

\newcommand{\dd}{\mathrm{d}}
\newcommand{\ee}{\mathrm{e}}
\newcommand{\ii}{\mathrm{i}}

\newcommand{\calP}{\mathcal{P}}
\newcommand{\calT}{\mathcal{T}}
\newcommand{\calU}{\mathcal{U}}
\newcommand{\calV}{\mathcal{V}}

\newcommand{\ovlU}{{\overline{U}}}

\newcommand{\ovlcalU}{{\overline{\mathcal{U}}}}
\newcommand{\ovlcalV}{{\overline{\mathcal{V}}}}

\newcommand{\eV}{\mathrm{eV}}

\newcommand{\plus}{{\mbox{{\bf{\tiny +}}}}}

\begin{document}

%
%
\title{Dirac Hamiltonian with Imaginary Mass and\\
Induced Helicity--Dependence by Indefinite Metric}

\newcommand{\addrMST}{Department of Physics,
Missouri University of Science and Technology,
Rolla, Missouri 65409-0640, USA}
\newcommand{\addrHD}{Institut f\"{u}r Theoretische Physik,
Philosophenweg 16, 69020 Heidelberg, Germany}

\author{U. D. Jentschura}
\affiliation{\addrMST}
\affiliation{\addrHD}

\begin{abstract}It is of general theoretical interest to 
investigate the properties of superluminal matter wave
equations for spin one-half particles.
One can either enforce superluminal 
propagation by an explicit substitution of the 
real mass term for an imaginary mass, 
or one can use a matrix representation of the 
imaginary unit that multiplies the mass term.
The latter leads to the tachyonic Dirac equation, 
while the equation obtained by the substitution 
$m \to \ii \, m$ in the Dirac equation 
is naturally referred to as the 
imaginary-mass Dirac equation.
Both the tachyonic as well as the imaginary-mass
Dirac Hamiltonians commute with the helicity operator.
Both Hamiltonians are pseudo-Hermitian and also possess
additional modified pseudo-Hermitian properties, leading to constraints
on the resonance eigenvalues. 
Here, by an explicit calculation, 
we show that specific sum rules over the
spectrum hold for the wave functions corresponding to the  well-defined real
energy eigenvalues and complex resonance and anti-resonance energies.
In the quantized imaginary-mass Dirac field,
one-particle states
of right-handed helicity acquire a negative norm 
(``indefinite metric'') and can be
excluded from the physical spectrum by a Gupta--Bleuler
type condition.
\end{abstract}

\pacs{95.85.Ry, 11.10.-z, 03.70.+k}

\maketitle

%
%
\section{Introduction and Overview}
\label{intro}

%
%
\subsection{Theory and Experiment}

The superluminal propagation of matter waves is a highly intriguing subject
which is not without controversy.  The subluminal (tardyonic) energy-momentum
relation $E^2 = \vec p^2 + m^2$ needs to be changed to the superluminal
(tachyonic) dispersion relation $E^2 = \vec p^2 - m^2$.  Recently, it has been
argued that the tachyonic Dirac equation~\cite{JeWu2011jpa,JeWu2012} provides for a
convenient framework for the description of tachyonic particles; in this
equation, the mass is multiplied by a matrix representation of the imaginary
unit. Here, starting from the Dirac Hamiltonian, we explore a Dirac equation
where the mass is explicitly multiplied by the imaginary unit and we find certain
fundamental relations for the corresponding spin-$1/2$ field theory.  We also
explore certain algebraic properties of modified Dirac theories with an
imaginary mass term, and pertaining consequences for the eigenvalue spectrum of
the imaginary-mass Dirac Hamiltonian.
The tachyonic formulation~\cite{BiDeSu1962,Fe1967,ArSu1968,DhSu1968,BiSu1969,%
Fe1977} of a fundamental field theory is the only one compatible 
with Lorentz invariance, and therefore, compatible
with special relativity. We exclusively use this concept
in the following and avoid any breaking of Lorentz 
invariance.

According to the summary overview presented in 
Ref.~\cite{LABneutrino}, low-energy experiments have determined the 
neutrino mass square to be in the range of a few $\eV^2$.
The best estimate for the neutrino mass square has 
been determined as negative in all 
experiments~\cite{RoEtAl1991,AsEtAl1994,StDe1995,AsEtAl1996,%
WeEtAl1999,LoEtAl1999,BeEtAl2008}, but the result has been 
consistent with a vanishing neutrino mass within 
experimental error bars.  
In direct measurements of the neutrino 
velocity~\cite{KaEtAl1979,AdEtAl2007,ICARUS2012},
the best estimate derived from experimental data
has been superluminal ($v > c$), 
but again, consistent with the hypothesis $v = c$ within
experimental error bars
(see also Ref.~\cite{LABneutrino} or Table~1 of Ref.~\cite{Ci1998}).
The OPERA collaboration~\cite{OPERA2011dracos}  has indicated a preliminary, revised
result of $(v-c)/c = (2.7 \pm 3.1 \mbox{(stat)} {}^{+3.8}_{-2.8} \mbox{(sys.)})
\times 10^{-6}$, which (just like all other available experimental 
results) neither excludes subluminal nor superluminal propagation.

The neutrino is generally regarded as the most 
prominent candidate for a superluminal particle in the low-energy 
domain~\cite{ChHaKo1985,ChKoPoGa1992,ChKo1994,Re1997,Ci1998},
However, the existence of conceivable superluminal particles
in hitherto unexplored kinematic regions cannot be excluded, either;
our study is of theoretical nature and not tied to a specific 
particle. It has recently been argued~\cite{JeWu2011jpa,JeWu2012} that the 
tachyonic theory of spin-$1/2$ particles is easier to implement
as compared to spinless particles, and we here continue
this line of thought by analyzing a theory where the 
imaginary mass is used explicitly in the Dirac equation,
rather than a matrix representation thereof.
The latter has been used 
in Refs.~\cite{ChHaKo1985,Ch2000,Ch2002a,JeWu2011jpa,JeWu2012}.
We use natural units with $\hbar = c = \epsilon_0 = 1$.

%
%
\subsection{Theoretical Foundations}

It is useful to recall that the subluminal (tardyonic) 
Dirac Hamiltonian $H_D$ reads
\begin{equation}
H_D = \vec \alpha \cdot \vec p + \beta \, m\,.
\end{equation}
Here, $\vec p$ is the momentum operator.
We use the Dirac matrices in the standard Dirac
representation ($\vec\alpha = \gamma^0 \, \vec\gamma$, 
and $\beta = \gamma^0$),
\begin{align}
\gamma^0 =& \; \left( \begin{array}{cc} \mathbbm{1}_{2\times 2} & 0 \\
0 & -\mathbbm{1}_{2\times 2} \\
\end{array} \right) \,,
\quad
\vec\gamma = \left( \begin{array}{cc} 0 & \vec\sigma \\ -\vec\sigma & 0  \\
\end{array} \right) \,.
\end{align}
The Hamiltonian $H_D$ 
can be modified into a Hamiltonian describing superluminal (tachyonic)
particles by the simple replacement $m \to \ii m$
(see Ref.~\cite{BaSh1974}),
leading to the imaginary-mass Dirac Hamiltonian
\begin{equation}
\label{Hprime}
H' = \vec \alpha \cdot \vec p + \ii \, \beta \, m \,.
\end{equation}
Alternatively, one can choose a matrix representation of the 
imaginary unit, and write the 
tachyonic Dirac Hamiltonian~\cite{ChHaKo1985,Ch2000,Ch2002a,JeWu2011jpa,JeWu2012} as
\begin{equation}
H_5 = \vec \alpha \cdot \vec p + \beta \, \gamma^5 m \,,
\end{equation}
with $(\beta \, \gamma^5)^2 = -\mathbbm{1}_{4 \times 4}$ and 
\begin{equation}
\gamma^5 = \left( \begin{array}{cc} 0 & \mathbbm{1}_{2\times 2} \\
\mathbbm{1}_{2\times 2} & 0  \\
\end{array} \right) \,.
\end{equation}
Both $H'$ and $H_5$ are pseudo-Hermitian,
which implies that eigenvalues are either real or
come in complex-conjugate pairs, $E$ and $E^*$.
Here, we also show that $H'$ and $H_5$ fulfill additional,
modified pseudo-Hermiticity conditions
(``quasi-pseudo-Hermiticity''), which allow us to further
conclude that if $E$ is a resonance eigenvalue, so is $-E^*$, 
and thus, the eigenvalues either come in (real) pairs 
$E$ and $-E^*$, or they occur in the rectangular complex configuration
$E$, $E^*$, $-E$, and $-E^*$. 
The quantization of the imaginary-mass Dirac theory 
naturally leads to helicity-dependent anticommutators.

We proceed as follows. In Sec.~\ref{algebra},
we derive a few algebraic properties of the Hamiltonians 
$H'$ and $H_5$ which determine the general properties of 
their spectra. The field theory defined 
by the Hamiltonian $H'$ is quantized
in Sec.~\ref{quantization}.
In Sec.~\ref{inversion}, we analyze the 
Hamiltonian $H''$ which is obtained from~\eqref{Hprime}
by the replacement $m \to -\widetilde m$.
The quantization of the imaginary-mass Dirac theory
is shown to yield rather interesting insight into
helicity-dependent anticommutators.
Conclusions are reserved for Sec.~\ref{conclu}.

%
%
\section{Algebraic Properties and Eigenvalues}
\label{algebra}

It is useful to derive a few algebraic 
properties of $H_5$ and $H'$ which determine the 
structure of the spectra of these Hamiltonians.
We explicitly refer to the coordinate-space 
representations ($\vec p = -\ii \, \hbar\vec \nabla$)
\begin{equation}
\label{HprimeC}
H'(\vec r) = -\ii \, \vec \alpha \cdot \vec \nabla + \ii \, \beta \, m \,,
\end{equation}
and 
\begin{equation}
\label{H5C}
H_5(\vec r) = -\ii \, \vec \alpha \cdot \vec \nabla + 
\beta \, \gamma^5 \, m \,.
\end{equation}
We use the following matrices,
\begin{equation}
\eta = \gamma^5 \,,
\qquad
\chi = \gamma^0 \,,
\qquad
\rho = \gamma^0 \, \gamma^5 \,.
\end{equation}
These fulfill $\eta^{-1} = \eta$, 
$\rho^{-1} = \gamma^5 \, \gamma^0 = -\rho$, and
$\chi^{-1} = \chi$. By an elementary calculation, 
we infer that
\begin{subequations} 
\label{HPPQP}
\begin{align} 
\label{HprimePSEUDO}
H'(\vec r) =& \; \eta \; H'^\plus(\vec r) \; \eta^{-1} \,,
\\[1.77ex]
\label{HprimeQPSEUDO}
H'(\vec r) =& \; -\chi\; H'^\plus(\vec r) \; \chi^{-1} \,,
\\[1.77ex]
\label{H5PSEUDO}
H_5(\vec r) =& \; \eta\; H_5^\plus(\vec r) \; \eta^{-1} \,,
\\[1.77ex]
\label{H5QPSEUDO}
H_5(\vec r) =& \; -\rho \; H_5^\plus(\vec r) \; \rho^{-1} \,,
\end{align} 
\end{subequations} 
where the superscript $\plus$ denotes the Hermitian adjoint.
The relations~\eqref{HprimePSEUDO} and~\eqref{H5PSEUDO}
imply the pseudo-Hermiticity of the Hamiltonians
$H'$ and $H_5$, respectively, in the sense of 
Refs.~\cite{Pa1943,BeBo1998,BeDu1999,%
BeBoMe1999,BeWe2001,BeBrJo2002,
Mo2002iii,Mo2003npb,JeSuZJ2009prl,JeSuZJ2010}.
As shown in Refs.~\cite{Pa1943,JeWu2011jpa},
for a pseudo-Hermitian Hamiltonian, 
if $E$ is a resonance eigenvalue, so is $E^*$. Indeed, it has 
been shown in Ref.~\cite{JeWu2012} that 
$H_5$ has both real eigenvalues (corresponding to 
plane-wave solutions of positive and negative energy),
and also resonances and anti-resonances whose resonance
energies are manifestly complex. The resonances correspond
to evanescent waves whose wavelength is too 
long to support superluminal propagation;
these waves therefore decay exponentially.

In comparison to the structure of 
Eqs.~\eqref{HprimePSEUDO} and~\eqref{H5PSEUDO},
the relations~\eqref{HprimeQPSEUDO} and~\eqref{H5QPSEUDO}
feature an additional minus sign.
They correspond to additional ``quasi-pseudo-Hermitian'' properties
of $H'$ and $H_5$.  These additional properties imply that if $E$ is 
a resonance eigenvalue, so is $-E^*$. This can be shown as follows.
Let $\psi$ be an eigenfunction of a general 
Hamiltonian $H'$ with eigenvalue $E$. 
Then, because the spectrum of the
Hermitian adjoint of an operator consists of the
complex-conjugate eigenvalues, there exists a
wavefunction $\phi$ with the property
\begin{equation} 
H'^\plus \, \phi = E^* \, \phi
\end{equation}
from which we infer that
\begin{equation} 
\left( \chi \, H'^\plus \, \chi'^{-1} \right) \;
\chi \, \phi = E^* \, \chi \, \phi \,,
\end{equation}
and so, in view of Eq.~\eqref{HprimeQPSEUDO},
we have $H' \, (\chi \, \phi) = -E^* \, (\chi \, \phi)$.
So, if $E$ is a resonance eigenvalue of $H'$, so is $-E^*$,
with a corresponding eigenvector $(\chi \, \phi)$.
The same property is implied for $H_5$ by Eq.~\eqref{H5QPSEUDO}.
If $E$ is real, then Eqs.~\eqref{HprimeQPSEUDO} and~\eqref{H5QPSEUDO}
imply that energy eigenvalues come in 
pairs $E$ and $-E$, whereas if they are manifestly complex,
then they exhibit a rectangular configuration
(in the complex plane)
consisting of $E$, $-E$, $E^*$ and $-E^*$.

%
%
\section{Quantization and Spin Sums}
\label{quantization}

First, we observe that both $H'$ and $H_5$ commute with the helicity 
operator,
\begin{equation}
\left[ \vec\Sigma \cdot \vec p, H' \right] =
\left[ \vec\Sigma \cdot \vec p, H_5 \right] = 0 \,,
\end{equation}
where
\begin{equation}
\vec\Sigma = \gamma^5 \, \gamma^0 \, \vec \gamma = 
\left( \begin{array}{cc} \vec\sigma & 0 \\
0 & \vec\sigma \\
\end{array} \right) \,.
\end{equation}
The quantization of the tachyonic theory 
defined by the Hamiltonian $H_5$ has been discussed
in Ref.~\cite{JeWu2012}.
Here, we are concerned with the Hamiltonian $H'$.
The corresponding covariant form the 
imaginary-mass Dirac equation reads as
\begin{equation}
\label{immassC}
(\ii \gamma^\mu \, \partial_\mu - \ii \, m) \, \psi(x) = 0 \,.
\end{equation}

Of course, it could be argued that the 
solutions of the imaginary-mass Dirac equation can be written down 
immediately, by simply replacing $m \to \ii \, m$
in the well-known bispinor solutions of the 
ordinary Dirac equation, as given in Chapter~2 of Ref.~\cite{ItZu1980}.
However, this procedure does not lead to compact formulas
when one tries to develop the formalism further. 
A brief, sketchy, illustrative remark is in order.
According to Eq.~(2.40) of Ref.~\cite{ItZu1980}, the 
spin sum over the positive-energy states of the 
tardyonic (ordinary) Dirac equation leads to the expression
\begin{equation}
\frac{1}{2m(m+E)} \, 
(\cancel{k} + m) \, \frac{1 + \gamma^0}{2} \,
(\cancel{k} + m) = \frac{\cancel{k} + m}{2 m} \,,
\end{equation}
where the latter term is the projector onto positive-energy states.
Here, $\cancel{k} = \gamma^\mu \, k_\mu$ is the Feynman dagger.
When replacing $m \to \ii \, m$ in the solution of the Dirac 
equation given in Eq.~(2.37) of Ref.~\cite{ItZu1980}, and 
performing the same spin sum over positive-energy solutions
of the form $\sum_\alpha u^{(\alpha)} \otimes \bar u^{(\alpha)}$
(using the notation of Ref.~\cite{ItZu1980}),
one has to replace $m \to \ii \, m$ for the spinors
and  $m \to -\ii \, m$ for the Dirac adjoint bispinors.
But then,
\begin{equation}
\frac{1}{2m(m+E)} \, 
(\cancel{k} + \ii \, m) \, \frac{1 + \gamma^0}{2} \,
(\cancel{k} - \ii \, m) \neq \frac{\cancel{k} + \ii m}{2 m} \,,
\end{equation}
which is not equal to a compact projector form,
as an elementary calculation shows.
By contrast, compact formulas for sums over spins can be obtained 
in the helicity basis, as shown in the following. 

For tachyonic particles, in analogy to the 
formalism developed in Ref.~\cite{JeWu2012},
it appears advantageous to 
use the helicity basis for the construction of the 
elementary solutions.
We recall that the eigenfunctions of the 
operator $\vec\sigma \cdot \vec k$ are given by
\begin{equation}
a_+(\vec k) = \left( \begin{array}{c} 
\cos\left(\frac{\theta}{2}\right) \\[1.77ex]
\sin\left(\frac{\theta}{2}\right) \, \ee^{\ii \, \varphi} \\
\end{array} \right) \,,
\quad
a_-(\vec k) = \left( \begin{array}{c} 
-\sin\left(\frac{\theta}{2}\right) \, \ee^{-\ii \, \varphi} \\[1.77ex]
\cos\left(\frac{\theta}{2}\right) \\
\end{array} \right) \,.
\end{equation}
where $\theta$ and $\varphi$ constitute the polar and azimuthal angles
of the wave vector $\vec k$,
with $(\vec \sigma \cdot \vec k) \, a_\pm(\vec k) =
\pm |\vec k|\,a_\pm(\vec k)$.
We also recall the normalized positive-energy 
chirality and helicity eigenspinors of the 
massless Dirac equation as follows
($C = \ii \, \gamma^2 \, \gamma^0$),
\begin{subequations}
\label{uv}
\begin{align}
u_+(\vec k) = & \;
\frac{1}{\sqrt{2}} 
\left( \begin{array}{c}
a_+(\vec k) \\[1.77ex]
a_+(\vec k) \\
\end{array} \right) \,,
\quad
u_-(\vec k) = 
\frac{1}{\sqrt{2}} 
\left( \begin{array}{c}
a_-(\vec k) \\[1.77ex]
-a_-(\vec k) \\
\end{array} \right) \,,
\\[1.77ex]
v_+(\vec k) = & \; C \, \overline u_-(\vec k)^{\rm T} =
\frac{1}{\sqrt{2}} 
\left( \begin{array}{c}
-a_+(\vec k) \\[1.77ex]
-a_+(\vec k) \\
\end{array} \right) 
= -u_+(\vec k)\,,
\\[1.77ex]
v_-(\vec k) =& \; C \, \overline u_+(\vec k)^{\rm T} =
\frac{1}{\sqrt{2}} 
\left( \begin{array}{c}
-a_-(\vec k) \\[1.77ex]
a_-(\vec k) \\
\end{array} \right) 
= -u_-(\vec k)\,.
\end{align}
\end{subequations}
Canonically, the subscripts $\pm$ of the $u$ and $v$ spinors 
correspond to the chirality (eigenvalue of $\gamma^5$), 
which (in the massless limit) 
is equal to helicity for positive-energy eigenstates,
and equal to the negative of the chirality for 
negative-energy eigenstates.
This is because the positive- and negative-energy 
solutions are multiplied by $\exp(\ii \vec k \cdot \vec r)$
and $\exp(-\ii \vec k \cdot \vec r)$, respectively
[see Eq.~\eqref{PsiPhi} below].
Using the relation
$(\cancel{k} + \ii \, m) \, (\cancel{k} - \ii \, m) = 
k^2 + m^2 = E^2 - \vec k^{\,2} + m^2$, we find
\begin{subequations}
\label{UU}
\begin{align}
U'_+(\vec k) = & \;
=
\left( \begin{array}{c}
\sqrt{\dfrac{E + \ii m}{2 \, |\vec k|}} \; a_+(\vec k) \\[2.33ex]
\sqrt{\dfrac{E - \ii m}{2 \, |\vec k|}} \; a_+(\vec k) \\
\end{array} \right) \,,
\\[1.77ex]
U'_-(\vec k) = & \;
\left( \begin{array}{c}
\sqrt{\dfrac{E + \ii m}{2 \, |\vec k|}} \; a_-(\vec k) \\[2.33ex]
-\sqrt{\dfrac{E - \ii m}{2 \, |\vec k|}} \; a_-(\vec k) \\
\end{array} \right)\,.
\end{align}
\end{subequations}
The massless limit $m \to 0$ ($E \to |\vec k|$)
is recovered as $U'_+(\vec k) \to u_+(\vec k)$ and
$U'_-(\vec k) \to u_-(\vec k)$.
The negative-energy eigenstates of the imaginary-mass Dirac equation
are given as
\begin{subequations}
\label{VV}
\begin{align}
V'_+(\vec k) = & \;
\left( \begin{array}{c}
-\sqrt{\dfrac{E - \ii m}{2 \, |\vec k|}} \; a_+(\vec k) \\[2.33ex]
-\sqrt{\dfrac{E + \ii m}{2 \, |\vec k|}} \; a_+(\vec k) \\
\end{array} \right) \,,
\\[1.77ex]
V'_-(\vec k) = & \;
\left( \begin{array}{c}
-\sqrt{\dfrac{E - \ii m}{2 \, |\vec k|}} \; a_-(\vec k) \\[2.33ex]
\sqrt{\dfrac{E + \ii m}{2 \, |\vec k|}} \; a_-(\vec k) \\
\end{array} \right)\,.
\end{align}
\end{subequations}
In the massless limit, the solutions
$v_+(\vec k)$ and $v_-(\vec k)$ are recovered,
$V'_+(\vec k) \to v_+(\vec k)$ and $V'_-(\vec k) \to v_-(\vec k)$. 
The states are normalized with respect to the condition
\begin{align}
U'^\plus_+(\vec k) \, U'_+(\vec k) =& \;
U'^\plus_-(\vec k) \, U'_-(\vec k) = 1 \,,
\nonumber\\[1.77ex]
V'^\plus_+(\vec k) \, V'_+(\vec k) =& \;
V'^\plus_-(\vec k) \, V'_-(\vec k) = 1 \,.
\end{align}
The positive- and negative-energy solutions
of the imaginary-mass Dirac equation are thus given as
\begin{subequations}
\label{PsiPhi}
\begin{align}
\Psi(x) =& \; U'_\pm(\vec k) \, \ee^{-\ii k \cdot x} \,,
\\[1.77ex]
\Phi(x) =& \; V'_\pm(\vec k) \, \ee^{\ii k \cdot x} \,,
\end{align}
\end{subequations}
Here, $\Psi$ is a solution for positive energy, and 
$\Phi$ constitutes a solution for negative energy.
All above formulas are valid for $|\vec k| \geq m$,
so that $E = \sqrt{\vec k^{2} - m^2}$ is real rather than complex.
For $|\vec k | < m$, one encounters resonances,
which complete the spectrum.
These are derived from Eqs.~\eqref{UU} and~\eqref{VV}
by the identification
\begin{subequations}
\begin{align}
E =& \; \pm \sqrt{\vec k^{\,2} - m^2 - \ii \, \epsilon} = 
\mp \, \ii \, \frac{\Gamma}{2} \,,
\\[1.77ex]
\Gamma =& \; 2 \, \sqrt{m^2 - \vec k^{\,2}}  \,, 
\qquad
| \vec k | < m \,.
\end{align}
\end{subequations}
The Dirac adjoint is
$\ovlU'_\sigma(\vec k) = U'^\plus_\sigma(\vec k) \, \gamma^0$.
By an elementary calculation, one shows that 
$\ovlU'_+(\vec k) \, U'_+(\vec k) = 
\ovlU'_-(\vec k) \, U'_-(\vec k) = 
\overline V'_+(\vec k) \, V'_+(\vec k) = 
\overline V'_-(\vec k) \, V'_-(\vec k) = 0$.
This can otherwise be seen as follows. One first realizes that
the adjoint equation of $\left( \cancel{k} - \ii \, m \right) \, U'_{\pm}(\vec k) = 0$
reads as
$U'_\pm(\vec k) \, \left( \cancel{k} + \ii \, m \right) = 0$,
and so
\begin{align}
\label{UpUp}
U'_\pm(\vec k) \, U'_\pm(\vec k) =& \;
\frac{1}{2 \ii m}  \,
U'_\pm(\vec k) \,
\left( \ii m + \ii m\right) \,
U'_\pm(\vec k) \nonumber\\
=& \;
\frac{1}{2 \ii m}  \,
U'_\pm(\vec k) \,
\left( -\cancel{k} + \cancel{k} \right) \, U'_\pm(\vec k) = 0 \,.
\end{align}
Here, we have used the adjoint equation for the
`first' $\ii \, m$ and the original form of the 
imaginary-mass Dirac equation for the `second' $\ii \, m$.

In analogy to Ref.~\cite{JeWu2012}, we define
the $\calU'$ and $\calV'$ bispinors in the following normalization,
\begin{subequations}
\label{covariant}
\begin{align}
\calU'_\sigma(\vec k) =& \; 
\left( \frac{|\vec k|}{m} \right)^{1/2} \, U'_\sigma(\vec k) \,,
\\[1.77ex]
\calV'_\sigma(\vec k) =& \; 
\left( \frac{|\vec k|}{m} \right)^{1/2} \, V'_\sigma(\vec k) \,.
\end{align}
\end{subequations}
Under charge conjugation, the spinors transform as
$C \overline \calU'_\pm(\vec k)^{\rm T} = -\calU'_\mp(\vec k)$
and $C \overline \calV'_\pm(\vec k)^{\rm T} = -\calV'_\mp(\vec k)$.
In analogy with Ref.~\cite{JeWu2012},
we write the field operator as
\begin{align}
\label{lines}
\psi(x) =& \;
\int \frac{\dd^3 k}{(2\pi)^3} \, 
\frac{m}{E} \sum_{\sigma = \pm} 
\left( b_\sigma(k) \, \calU'_\sigma(\vec k) \, 
\ee^{-\ii \, k \cdot x} \right.
\nonumber\\[1.77ex]
& \; \left. 
+ b_\sigma(-k) \, \calV'_\sigma(\vec k) \, 
\ee^{\ii \, k \cdot x} \right) \,,
\nonumber\\[1.77ex]
k =& \; (E, \vec k)\,,
\qquad E = E_{\vec k} =  \sqrt{\vec k^2 - m^2 - \ii \, \epsilon} \,.
\end{align}
Note that the $\ii \epsilon$ prescription selects the resonances 
(as analytic continuations of the positive-energy solutions)
and antiresonances (as analytic continuations of the negative-energy solutions).
This ensures that the waves are evanescent in their 
respective propagation direction in time.
The second term in the sum in Eq.~\eqref{lines}
describes the absorption of a negative-energy 
tachyonic particle that propagates backward in time;
this process is of course equivalent to the emission of a positive-energy 
antiparticle propagating forward in time by the Feinberg--Sudarshan 
reinterpretation principle, as explained in Ref.~\cite{JeWu2012}.
Thus,
\begin{align}
\label{fieldop}
\psi(x) =& \;
\int \frac{\dd^3 k}{(2\pi)^3} \, 
\frac{m}{E} \sum_{\sigma = \pm} 
\left( b_\sigma(k) \, \calU'_\sigma(\vec k) \, 
\ee^{-\ii \, k \cdot x} \right.
\nonumber\\[1.77ex]
& \; \left. 
+ d^\plus_\sigma(k) \, \calV'_\sigma(\vec k) \, 
\ee^{\ii \, k \cdot x} \right)  \,,
\end{align}
where $d^\plus_\sigma$ creates antiparticles.
For the imaginary-mass formalism, we postulate the same anticommutators as
in Ref.~\cite{JeWu2012} for the tachyonic Dirac equation,
\begin{subequations}
\label{anticom}
\begin{align}
\left\{ b_\sigma(k) , b_{\rho}(k') \right\} = & \;
\left\{ b^\plus_\sigma(k) , b^\plus_{\rho}(k') \right\} = 0 \,,
\\[1.77ex]
\left\{ d_\sigma(k) , d_{\rho}(k') \right\} = & \;
\left\{ d^\plus_\sigma(k) , d^\plus_{\rho}(k') \right\} = 0 \,.
\end{align}
\end{subequations}
The nonvanishing anticommutators read as follows,
\begin{subequations}
\begin{align}
\left\{ b_\sigma(k) , b^\plus_{\rho}(k') \right\} = & \; 
(-\sigma) \, 
(2 \pi)^3 \, \frac{E}{m} \delta^3(\vec k - \vec k') \, 
\delta_{\sigma\rho}\,,
\\[1.77ex]
\left\{ d_\sigma(k) , d^\plus_{\rho}(k') \right\} = & \; 
(-\sigma) \, 
(2 \pi)^3 \, \frac{E}{m} \delta^3(\vec k - \vec k') \, 
\delta_{\sigma\rho}\,.
\end{align}
\end{subequations}
The $\sigma$-dependent
anticommutator implies that the norm of the right-handed helicity 
(positive chirality) neutrino one-particle state is negative,
and that the right-handed helicity particle 
state has negative norm and can be excluded from the 
physical spectrum if one imposes a Gupta--Bleuler type condition
(according to Chapter~3 of Ref.~\cite{ItZu1980}).
Likewise, antiparticles described by the imaginary-mass formalism
can only exist in the right-handed helicity state.
The matrix-valued tachyonic field anticommutator reads as
\begin{align}
& \{ \psi(x), \overline\psi(y) \} = 
\left< 0 \left| \{ \psi(x), 
\overline\psi(y) \} \right| 0 \right> 
\nonumber\\[1.77ex]
& = \int \frac{\dd^3 k}{(2 \pi)^3} 
\frac{m}{E} \,
\sum_{\sigma = \pm} \left(
\ee^{-\ii k \cdot (x-y)} \,
\left(-\sigma\right) \, \calU'_{\sigma}(\vec k) \otimes
\overline \calU'_{\sigma}(\vec k) \right.
\nonumber\\[1.77ex]
& \qquad \left. +
\ee^{\ii k \cdot (x-y)} \,
\left(-\sigma\right) \, \calV'_{\sigma}(\vec k) \otimes
\overline \calV'_{\sigma}(\vec k) \right) \,,
\end{align}
where $\sigma$ is the helicity (for positive-energy states)
and the negative of the helicity (for negative-energy states) 
and $\otimes$ is the tensor product in bispinor space.
The following two relations
\begin{subequations}
\label{tensor}
\begin{align}
\label{tensora}
\sum_\sigma (-\sigma) \; \calU'_\sigma(\vec k) \otimes
\ovlcalU'_\sigma(\vec k) \,\gamma^5 =& \; 
\frac{\cancel{k} + \ii \,  m}{2 m} \,,
\\[1.77ex]
\label{tensorb}
\sum_\sigma (-\sigma) \; \calV'_\sigma(\vec k) \otimes
\ovlcalV'_\sigma(\vec k) \,\gamma^5 =& \; 
\frac{\cancel{k} - \ii \, m}{2 m} \,,
\end{align}
\end{subequations}
are analogous to those found for the bispinor solutions of the 
tachyonic Dirac described in Ref.~\cite{JeWu2011jpa}.
Note that the factors $(-\sigma)$ in these equations are due to 
the quantization conditions~\eqref{anticom}.
Using Eq.~\eqref{tensor}, we can derive the compact result,
\begin{equation}
\label{anticom2}
\{ \psi(x), \overline\psi(y) \} \, \gamma^5 = 
\left( \ii \, \cancel{\partial} + \ii \, m \right)
\ii \, \Delta(x - y) \,,
\end{equation}
where $\Delta(x-y)$ is the distribution encountered in 
Eqs.~(3.55) and (3.56) of Ref.~\cite{ItZu1980},
\begin{equation}
\ii \, \Delta(x-y) =
\int \frac{\dd^3 k}{(2 \pi)^3} \, 
\frac{1}{2 E} \, \left( \ee^{-\ii k \cdot (x-y)} -
\ee^{\ii k \cdot (x-y)} \right) \,.
\end{equation}
The equal-time anticommutator of the fields thus reads as
$\left. \{ \psi(x), \overline\psi(y) \} \, \gamma^5 
\right|_{x_0 = y_0} 
= \gamma^0  \, \delta^3(\vec r - \vec s)$,
with the full, unfiltered Dirac-$\delta$ function
and $x =(t, \vec r)$ as well as $y= (t, \vec s)$
and the time $x_0 = y_0 = t$. Furthermore, with the help of 
Eqs.~\eqref{fieldop} and~\eqref{tensor},
one obtains the propagator $S'$ (time-ordered product),
\begin{subequations}
\label{ST}
\begin{align}
\label{STa}
& \left< 0 \left| T \, \psi(x) \, 
\overline{\psi}(y) \gamma^5 \right| 0 \right> = 
\ii \, S'(x - y) \,,
\\[1.77ex]
\label{STb}
& S'(x - y) = 
\int \frac{\dd^4 k}{(2 \pi)^4} \, 
\ee^{-\ii k \cdot (x-y)} \,
\frac{\cancel{k} + \ii \, m}{k^2 + m^2 + \ii \, \epsilon} \,.
\end{align}
\end{subequations}
The chirality projectors are invariant under multiplication by $\gamma^5$,
in view of the relation $\gamma^5(1 \pm \gamma^5)/2=\pm (1 \pm \gamma^5)/2$.

For consistency reasons, 
the imaginary-mass Dirac propagator should be 
connected with a Green function, 
%
\begin{equation}
\label{greenf}
S' = \gamma^0 \, \frac{1}{E - H'} \,,
\end{equation}
where $E$ is the energy argument of the Green 
function and $H'$ is the imaginary-mass Dirac Hamiltonian.
In momentum space, we can replace $H'$
by $\vec\alpha \cdot \vec k + \ii \, \beta \, m$.
An elementary calculation then shows that
\begin{equation}
S'(k) = \frac{1}{\cancel{k} - \ii \, m} = 
 \frac{\cancel{k} + \ii  \, m}{k^2 + m^2} \,.
\end{equation}
Introducing the $\ii \epsilon$ prescription
as before, we find that 
\begin{equation}
\label{ST2}
S'(k) = \frac{1}{\cancel{k} - \ii \, (m + \ii\,\epsilon)} =
\frac{\cancel{k} + \ii \, m}{k^2 + m^2 + \ii \, \epsilon} \,.
\end{equation}
%

Having determined the propagator, let us briefly
comment on the non-invariance of the 
imaginary-mass Dirac Hamiltonian under time reversal.
Indeed, time reversal exchanges the in- and out-states of a process.
In the calculation of a cross section, 
one has to square an invariant amplitude, 
which also exchanges in- and out-states, and 
leads to the occurrence of a propagator of the form 
\begin{equation}
\gamma^0 \, S'^\plus(k) \, \gamma^0 = 
\frac{\cancel{k} - \ii \, m}{k^2 + m^2 + \ii \, \epsilon} \,,
\end{equation}
which is obtained from~\eqref{ST2} under the replacement
$\ii \, m \to -\ii \, m$. In the time-reversed
Hamiltonian, according to Ref.~\cite{BeBrReRe2004},
the same replacement takes place. So, the non-invariance under 
time reversal
of the imaginary-mass Dirac equation does not necessarily lead to an 
inconsistent formalism within field theory.

%
%
\section{Inversion of the Mass Term}
\label{inversion}

It is instructive to consider the Hamiltonian 
which is obtained from the imaginary-mass Dirac Hamiltonian
in Eq.~\eqref{Hprime} by the replacement $m \to - \widetilde m$,
which amounts to an inversion of the sign of the mass term,
\begin{equation}
\label{Hdprime}
H'' = \vec \alpha \cdot \vec p - \ii \, \beta \, \widetilde m \,,
\end{equation}
A preliminary remark is in order.
Within $\calP \calT$ symmetric quantum 
mechanics~\cite{BeBo1998,BeDu1999,BeBoMe1999,BeWe2001,BeBrJo2002,
Mo2002iii,Mo2003npb,%
JeSuZJ2009prl,JeSuZJ2010}, the 
one-dimensional quantum mechanical Hamiltonians
$h' = -\partial_x^2 + \ii \, |G| \, x^3$ and 
$h'' = -\partial_x^2 - \ii \, |G| \, x^3$ 
(with $x$ being the coordinate)
have been used as paradigmatic examples of an
anharmonic (cubic) oscillator with imaginary coupling $\ii \, |G|$.
The Hamiltonians $h'$ and $h''$ have the same 
spectrum~\cite{BeDu1999,BeWe2001,JeSuZJ2009prl,JeSuZJ2010},
and moreover, the eigenvalues can be shown to 
be analytic functions in the complex $G$ plane 
where $\ii \, G = \sqrt{g}$, and the $g$ plane
has a branch cut along the negative real axis.

As is to be expected, the Hamiltonians $H'$ and $H''$ 
have the same spectrum, because $H''$ fulfills the same 
algebraic relations~\eqref{HprimePSEUDO} and~\eqref{HprimeQPSEUDO}
as $H'$. Moreover, the plane-wave eigenstates of 
$H''$ are solutions of the covariant equation
\begin{equation}
\label{immassCdprime}
(\ii \gamma^\mu \, \partial_\mu + \ii \, \widetilde m) \, \psi(x) = 0 \,,
\end{equation}
where
\begin{equation}
\Psi(x) = U''_\pm(\vec k) \, \ee^{-\ii k \cdot x} \,,
\qquad
\Phi(x) = V''_\pm(\vec k) \, \ee^{\ii k \cdot x} 
\end{equation}
for positive-energy and negative-energy states, respectively.
We find
\begin{subequations}
\label{UUdprime}
\begin{align}
U''_+(\vec k) = & \;
=
\left( \begin{array}{c}
\sqrt{\dfrac{E - \ii \widetilde m}{2 \, |\vec k|}} \; a_+(\vec k) \\[2.33ex]
\sqrt{\dfrac{E + \ii \widetilde m}{2 \, |\vec k|}} \; a_+(\vec k) \\
\end{array} \right) \,,
\\[1.77ex]
U''_-(\vec k) = & \;
\left( \begin{array}{c}
\sqrt{\dfrac{E - \ii \widetilde m}{2 \, |\vec k|}} \; a_-(\vec k) \\[2.33ex]
-\sqrt{\dfrac{E + \ii \widetilde m}{2 \, |\vec k|}} \; a_-(\vec k) \\
\end{array} \right)\,.
\end{align}
\end{subequations}
The negative-energy eigenstates are given as
\begin{subequations}
\label{VVdprime}
\begin{align}
V''_+(\vec k) = & \;
\left( \begin{array}{c}
-\sqrt{\dfrac{E + \ii \widetilde m}{2 \, |\vec k|}} \; a_+(\vec k) \\[2.33ex]
-\sqrt{\dfrac{E - \ii \widetilde m}{2 \, |\vec k|}} \; a_+(\vec k) \\
\end{array} \right)  \,,
\\[1.77ex]
V''_-(\vec k) = & \;
\left( \begin{array}{c}
-\sqrt{\dfrac{E + \ii \widetilde m}{2 \, |\vec k|}} \; a_-(\vec k) \\[2.33ex]
\sqrt{\dfrac{E - \ii \widetilde m}{2 \, |\vec k|}} \; a_-(\vec k) \\
\end{array} \right)\,.
\end{align}
\end{subequations}
The states are normalized with respect to the condition
\begin{subequations}
\begin{align}
U''^\plus_+(\vec k) \, U''_+(\vec k) =& \;
U''^\plus_-(\vec k) \, U''_-(\vec k) = 1 \,,
\\[1.77ex]
V''^\plus_+(\vec k) \, V''_+(\vec k) =& \;
V''^\plus_-(\vec k) \, V''_-(\vec k) = 1 \,.
\end{align}
\end{subequations}
We normalize the $\calU''$ and $\calV'$ bispinors according to 
\begin{subequations}
\begin{align}
\calU''_\sigma(\vec k) =& \; 
\left( \frac{|\vec k|}{m} \right)^{1/2} \, U''_\sigma(\vec k) \,,
\\[1.77ex]
\calV''_\sigma(\vec k) =& \; 
\left( \frac{|\vec k|}{m} \right)^{1/2} \, V''_\sigma(\vec k) \,,
\end{align}
\end{subequations}
and obtain the following two relations,
\begin{subequations}
\label{itensor}
\begin{align}
\label{itensora}
\sum_\sigma (-\sigma) \; \calU''_\sigma(\vec k) \otimes
\ovlcalU''_\sigma(\vec k) \,\gamma^5 =& \; 
\frac{\cancel{k} - \ii \,  \widetilde m}{2 \, \widetilde m} \,,
\\[1.77ex]
\label{itensorb}
\sum_\sigma (-\sigma) \; \calV''_\sigma(\vec k) \otimes
\ovlcalV''_\sigma(\vec k) \,\gamma^5 =& \; 
\frac{\cancel{k} + \ii \, \widetilde m}{2 \, \widetilde m} \,.
\end{align}
\end{subequations}
These are the analogues of Eqs.~\eqref{tensora} and~\eqref{tensorb}
and differ from Eqs.~\eqref{tensora} and~\eqref{tensorb}
by the replacement $m \to -\widetilde m$ in the numerator.
However, in the denominator no change takes place, 
because the denominator is obtained 
as $\sqrt{m^2} \to \sqrt{\widetilde m^2} = \widetilde m$.
The field operator is
\begin{align}
\psi(x) =& \;
\int \frac{\dd^3 k}{(2\pi)^3} \, 
\frac{\widetilde m}{E'} \sum_{\sigma = \pm} 
\left[ b'_\sigma(k) \, \calU''_\sigma(\vec k) \, 
\ee^{-\ii \, k \cdot x} \right.
\nonumber\\[1.77ex]
& \; \left. 
+ d'^\plus_\sigma(k) \, \calV''_\sigma(\vec k) \, 
\ee^{\ii \, k \cdot x} \right]  \,,
\end{align}
with $E' = \sqrt{\vec k^2 - \widetilde m^2}$,
and with an obvious identification of the 
field operators according to Eq.~\eqref{fieldop}.
The nonvanishing anticommutators read as follows,
\begin{subequations}
\begin{align}
\left\{ b'_\sigma(k) , b'^\plus_{\rho}(k') \right\} = & \; 
(-\sigma) \, 
(2 \pi)^3 \, \frac{E'}{\widetilde m} \delta^3(\vec k - \vec k') \, 
\delta_{\sigma\rho}\,,
\\[1.77ex]
\left\{ d'_\sigma(k) , d'^\plus_{\rho}(k') \right\} = & \; 
(-\sigma) \, 
(2 \pi)^3 \, \frac{E'}{\widetilde m} \delta^3(\vec k - \vec k') \, 
\delta_{\sigma\rho}\,.
\end{align}
\end{subequations}
These imply that the inversion of the mass term 
does not change the fact that again, {\em right-handed particle}
and {\em left-handed antiparticle} states acquire a 
negative norm. It is very instructive to clarify 
{\em by an explicit, detailed calculation} that the 
inversion of the mass term does not change the pattern 
by which helicity components are suppressed
for particle and antiparticle states.

%
%
\section{Conclusions}
\label{conclu}

In the current work, we investigate the relativistic
(tachyonic) quantum theory defined by the Hamiltonian
$H' = \vec \alpha \cdot \vec p + \ii \, \beta \, m$,
which is obtained from the ordinary Dirac Hamiltonian 
by the simple replacement $m \to \ii \, m$,
In Sec.~\ref{algebra},
we show that the Hamiltonian $H'$ is pseudo-Hermitian 
and has an additional quasi-pseudo-Hermitian
property given in Eq.~\eqref{HprimeQPSEUDO}.
Eigenvalues come in a specific structure in the complex plane.
Namely, if $E$ is a resonance eigenvalue, so is $E^*$, $-E$, and $-E^*$.
This pattern is manifest in the spectrum calculated for 
the tachyonic Dirac Hamiltonian 
$H_5$ in Ref.~\cite{JeWu2012} and in the 
spectrum of $H'$ calculated here.
Plane-wave solutions of the imaginary-mass Dirac equation 
are given in Eqs.~\eqref{UU} and~\eqref{VV}.

In Sec.~\ref{quantization},
we complement recent work on the tachyonic Dirac
Hamiltonian~\cite{JeWu2012} and discuss 
the quantization of the spin
one-half theory defined by the imaginary-mass Dirac Hamiltonian.
We find helicity-dependent anticommutators as given in 
Eq.~\eqref{anticom}.
For both the imaginary-mass as well as the tachyonic Dirac Hamiltonian,
the one-particle states of right-handed helicity acquire a negative norm and
can be excluded from the physical spectrum by a Gupta-Bleuler type condition.
Likewise, antiparticle states of left-handed helicity are excluded from the
physical spectrum. Compact representations are found for the spin
sums~\eqref{tensor} which enter the field anticommutator and the propagator.
In Sec.~\ref{inversion}, we find that an inversion 
of the mass term does not change the fact that
only left-handed helicity is allowed for particles
described by a tachyonic generalization of the Dirac equation,
and only right-handed helicity for antiparticles.

Obviously, the left-handedness of particle states
and the right-handedness of antiparticles states
imply that both the tachyonic Dirac equation as well as
the imaginary-mass Dirac equation represent candidates for the 
description of neutrinos, if improved experimental 
techniques~\cite{BiEtAl1987JETPL,RoEtAl1991,AdEtAl2007,OPERA2011v4,ICARUS2012}
finally allow us to decide if neutrinos propagate at 
superluminal or subluminal speeds, which would 
amount to deciding whether the neutrino mass square is 
positive or negative~\cite{RoEtAl1991,AsEtAl1994,StDe1995,AsEtAl1996,%
WeEtAl1999,LoEtAl1999,BeEtAl2008,LABneutrino}.

%
%
\section*{Acknowledgments}

This work was supported by the NSF and by the
National Institute of Standards and Technology
(precision measurement grant).
The author acknowledges helpful conversations with B.~J.~Wundt and
I.~N\'{a}ndori.

\end{document}